%% file: main.tex
\documentclass[conference]{IEEEtran}
\IEEEoverridecommandlockouts
\usepackage{cite}
\usepackage{url}
\usepackage[T1]{fontenc}
\usepackage{amsmath}
\usepackage{amssymb}
\usepackage{amsfonts}
\usepackage{graphicx}
\usepackage{xcolor}
\usepackage{xspace}
\usepackage{mathtools}
\usepackage{subfigure}
\usepackage{algorithm}
\usepackage{algorithmicx}
\usepackage{algpseudocode}
\usepackage{balance}
\usepackage{tikz}
\usepackage{pgfplots}
\usetikzlibrary{positioning}

\algrenewcommand\alglinenumber[1]{\scriptsize #1:}

\def\BibTeX{{\rm B\kern-.05em{\sc i\kern-.025em b}\kern-.08em
    T\kern-.1667em\lower.7ex\hbox{E}\kern-.125emX}}
    
\newcommand{\system}{{ParBlockchain}\xspace}
\newcommand{\orexp}{{OXII}\xspace}
\newcommand{\exe}{executor\xspace}
\newcommand{\exes}{executors\xspace}
\newcommand{\Exes}{Executors\xspace}
\newcommand{\ord}{orderer\xspace}
\newcommand{\ords}{orderers\xspace}
\newcommand{\Ords}{Orderers\xspace}
\newcommand{\REQ}{REQUEST\xspace}
\newcommand{\req}{request\xspace}
\newcommand{\NEWB}{NEWBLOCK\xspace}
\newcommand{\newb}{new block\xspace}
\newcommand{\CMT}{COMMIT\xspace}
\newcommand{\cmt}{commit\xspace}

\def\caT{{\cal T}}
\def\caE{{\cal E}}

\newenvironment{deff}{\medskip\noindent{\bf Definition:}~}{\medskip}

\begin{document}

\title{\system: Leveraging Transaction Parallelism in Permissioned Blockchain Systems}

\author{\IEEEauthorblockN{Mohammad Javad Amiri, Divyakant Agrawal, Amr El Abbadi}
\IEEEauthorblockA{University of California Santa Barbara}
Santa Barbara, California \\
\{amiri, agrawal, amr\}@cs.ucsb.edu}

\maketitle
\IEEEpeerreviewmaketitle

\input{sec_abstract}
\input{sec_intro}
\input{sec_background}
\input{sec_model}
\input{sec_algorithm}
\input{sec_exp}

\input{sec_related}
\input{sec_conc}
\balance

\bibliographystyle{plain}
\bibliography{main}

\end{document}

%% file: sec_abstract.tex
\begin{abstract}
Many existing blockchains
do not adequately address all the characteristics of
distributed system applications and
suffer from serious architectural limitations resulting
in performance and confidentiality issues.
While recent permissioned blockchain systems, have tried to 
overcome these limitations,
their focus has mainly been on workloads with no-contention, i.e.,
no conflicting transactions.
In this paper, we introduce {\em \orexp},
a new paradigm for permissioned blockchains
to support distributed applications that execute concurrently.
\orexp is designed for workloads with (different degrees of) contention.
We then present {\em \system}, a permissioned blockchain 
designed specifically in the \orexp paradigm.
The evaluation of
\system using a series of benchmarks
reveals that its performance in workloads with any degree of contention
is better than the state of the art permissioned blockchain systems.
\end{abstract}


%% file: sec_intro.tex
\section{Introduction}\label{sec:intro}

A blockchain is a distributed data structure for recording transactions
maintained by many nodes without a central authority \cite{cachin2017blockchain}.
In a blockchain, nodes agree on their shared states
across a large network of {\em untrusted} participants.
Blockchain was originally devised for Bitcoin cryptocurrency \cite{nakamoto2008bitcoin},
however, recent systems focus on its unique features such as
transparency, provenance, fault-tolerant, and authenticity
to support a wide range of distributed applications.
Bitcoin and other cryptocurrencies are {\em permissionless} blockchains.
In a permissionless blockchain, the network is public, and anyone can participate without a specific identity.
Many other distributed applications such as supply chain management \cite{korpela2017digital} and healthcare \cite{azaria2016medrec},
on the other hand, are deployed on {\em permissioned} blockchains
consisting of a set of known, identified nodes that still do not fully trust each other.

Distributed applications have different characteristics that need to be addressed by permissioned blockchain systems.
Such applications require high performance in terms of throughput and latency, e.g.,
a financial application needs to process tens of thousands of requests every second with very low latency.
Distributed applications might also have workloads with high-contention, i.e.,
conflicting transactions.
Under these workloads, several transactions simultaneously perform conflicting operations on
a few popular records.
These conflicting transactions might belong to a single application
or even a set of applications using a shared datastore.
While the sequential execution of transactions prevents any possible inconsistency,
it adversely impacts performance and scalability.
In addition, confidentiality of data is required in many applications.
In blockchain, the logic of each application can be written as a {\em smart contract},
as exemplified by Ethereum \cite{ethereum17}.
A smart contract is a computer program that self-executes once it is established and deployed.
Since smart contracts include the logic of applications, it might be desired to
restrict access to such contracts.
Cryptographic techniques are used to achieve confidentiality, however
the considerable overhead of such techniques makes them impractical \cite{androulaki2018hyperledger}.

Existing permissioned blockchains,
e.g., Tendermint \cite{kwon2014tendermint} and Multichain \cite{greenspan2015multichain},
mostly employ an {\em order-execute} paradigm where
nodes agree on a total order of the blocks of transactions using a consensus protocol and then
the transactions are executed in the same order on all nodes sequentially.
Such a paradigm suffers from performance issues because of
the sequential execution of transactions on all nodes, and also
confidentiality issues since every node access every smart contract.
Hyperledger Fabric \cite{androulaki2018hyperledger}, on the other hand,
presents a new paradigm for permissioned blockchains
by switching the order of the execution and ordering phases.
In Hyperledger Fabric, transactions of different applications are first executed in parallel and then
an ordering service consisting of a set of nodes uses a consensus protocol to establish
agreement on a total order of all transactions.
Fabric addresses the confidentiality issues by restricting accesses to smart contracts and
improves performance by executing transactions in parallel.
However, in the presence of any contention in the workload,
it has to disregard the effects of conflicting transactions which
negatively impacts the performance of the blockchain.

In this paper, we present {\em \orexp}: an order-execute paradigm for permissioned blockchains.
\orexp is mainly introduced to support distributed applications processing workloads with {\em some degree of contention}.
\orexp consists of {\em \ord} and {\em agent} nodes.
\Ords establish agreement on the order of the transactions of different applications,
construct the blocks of transactions, and
generate a {\em dependency graph} for the transactions within a block.
A dependency graph, on the one hand, gives a partial order based on the conflicts between
transactions, and,
on the other hand, enables higher concurrency by allowing
the parallel execution of non-conflicting transactions.
A group of agents of each application called {\em \exes}
are then responsible for executing the transactions of that application.

We then present {\em \system}, a permissioned blockchain system
designed specifically in the \orexp paradigm.
\system processes transactions in the ordering and execution phases.
In the ordering phase, transactions are ordered in a dependency graph and put in blocks.
In the execution phase, the \exes of each application execute
the transactions of the corresponding application
following the dependency graph.
As long as the partial order of transactions in the dependency graph is preserved,
the transactions of different applications
can be executed in parallel.

A key contribution of this paper is to show how workloads with conflicting transactions
can be handled efficiently by a blockchain system without
rolling back (aborting) the processed transactions or executing all transactions sequentially.
This paper makes the following contributions:

\begin{itemize}
\item {\em \orexp},
a new paradigm for permissioned blockchains
to support distributed applications that execute concurrently.
\orexp uses a dependency graph based concurrency control technique
to detect possible conflicts between transactions and to ensure the valid execution of transactions
while still allowing non-conflicting transactions to be executed in parallel.

\item {\em \system}, a permissioned blockchain system
designed specifically in the \orexp paradigm.
The experiments show that workloads with any degree of contention 
will benefit from \system.

\end{itemize}

The rest of this paper is organized as follows.
Section~\ref{sec:back} briefly describes current blockchain paradigms and their limitations.
The \orexp paradigm is introduced in Section~\ref{sec:model}.
Section~\ref{sec:system} presents \system, a permissioned blockchain system
designed specifically in the \orexp paradigm.
Section~\ref{sec:eval} shows the performance evaluation.
Section~\ref{sec:related} presents related work, and
Section~\ref{sec:conc} concludes the paper.

%% file: sec_background.tex
\section{Background}\label{sec:back}

A blockchain is a distributed data structure for recording transactions
maintained by many nodes without a central authority \cite{cachin2017blockchain}.
A blockchain replicates data
over nodes using State Machine Replication (SMR).
State machine replication is a technique for implementing a fault-tolerant service
by replicating servers \cite{lamport1978time}.
In the state machine replication model replicas agree on
an ordering of incoming requests and then execute the
requests in the same order.
State machine replication approaches have been used in different synchronous and asynchronous networks to tolerate
crash, malicious, or both failures. 
In a crash failure model, replicas 
may fail by stopping, and may restart, however, they may not collude, lie, or
otherwise, attempt to subvert the protocol.
In contrast, in a Byzantine failure model, faulty nodes may exhibit arbitrary,
potentially malicious, behavior.

Blockchains use asynchronous fault-tolerant protocols to establish consensus.
Since the nodes in a blockchain could behave maliciously,
blockchains mainly use Byzantine fault-tolerant protocols to reach consensus.

In general, "ordering" and "execution" are the two main tasks of any fault-tolerant system.
Fault-tolerant protocols mainly follow an {\em order-execute} paradigm where
the network first, orders transactions and then
executes them in the same order on all nodes sequentially.

  \begin{figure*}[t]
  \centering
  \subfigure[Order-Execute Paradigm (Permissionless)]{
  \label{fig:oe}
  \includegraphics[width=0.31\linewidth]{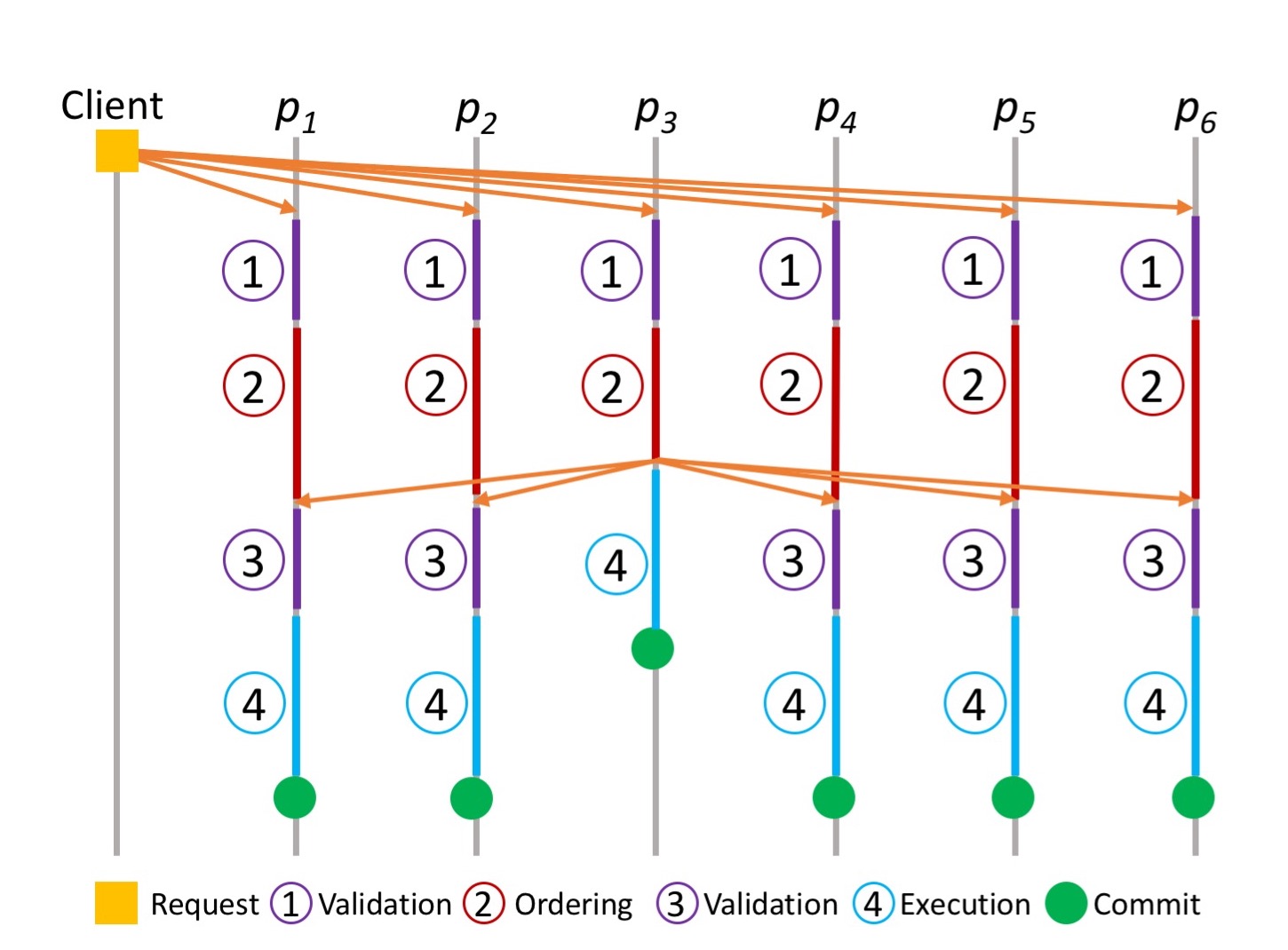}}
  \hfill
  \subfigure[Order-Execute Paradigm (Permissioned)]{
  \label{fig:oe-perm}
  \includegraphics[width=0.31\linewidth]{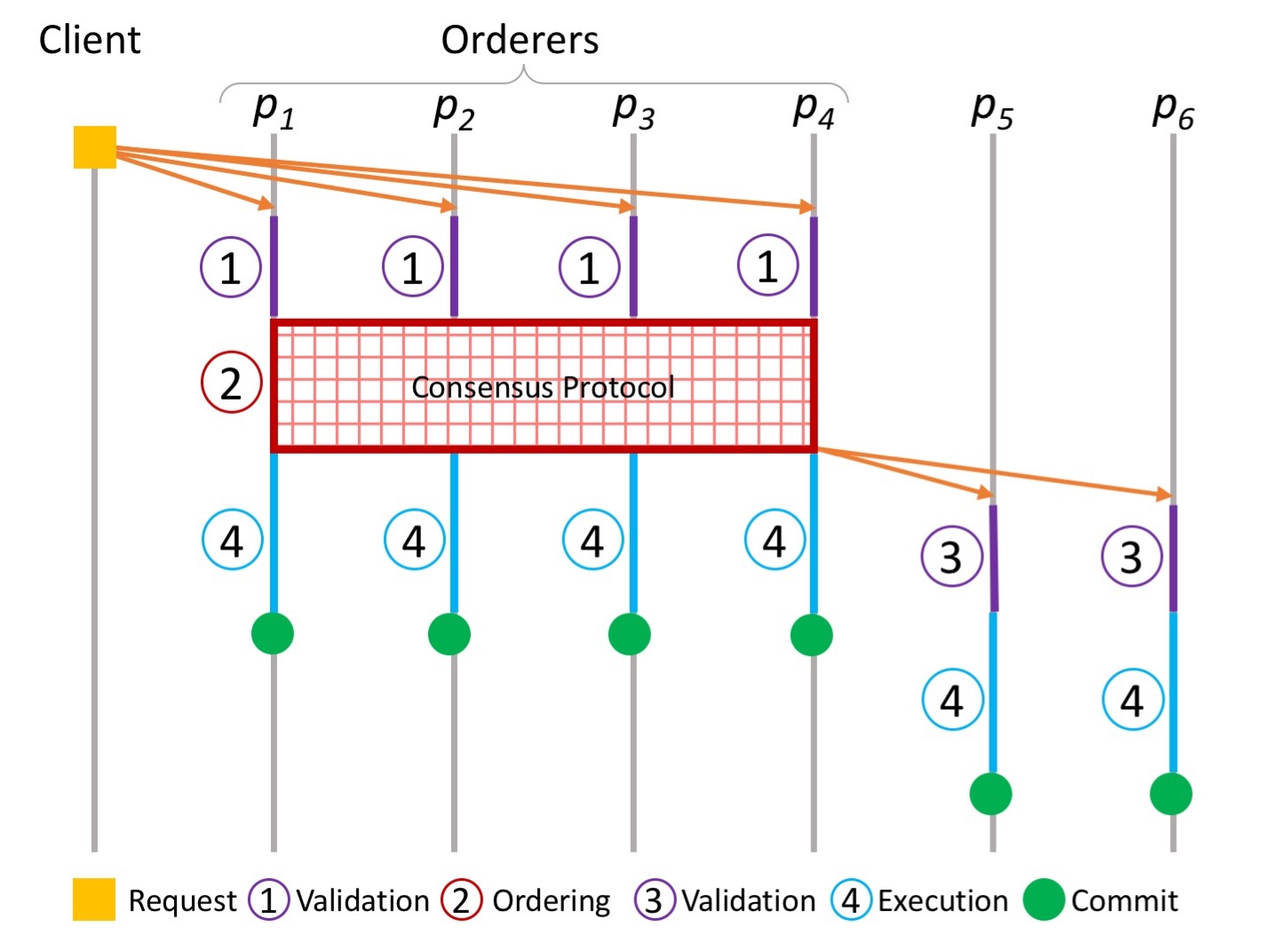}}
  \hfill
  \subfigure[Execute-Order Paradigm (Permissioned)]{
  \label{fig:eo}
  \includegraphics[width=0.31\linewidth]{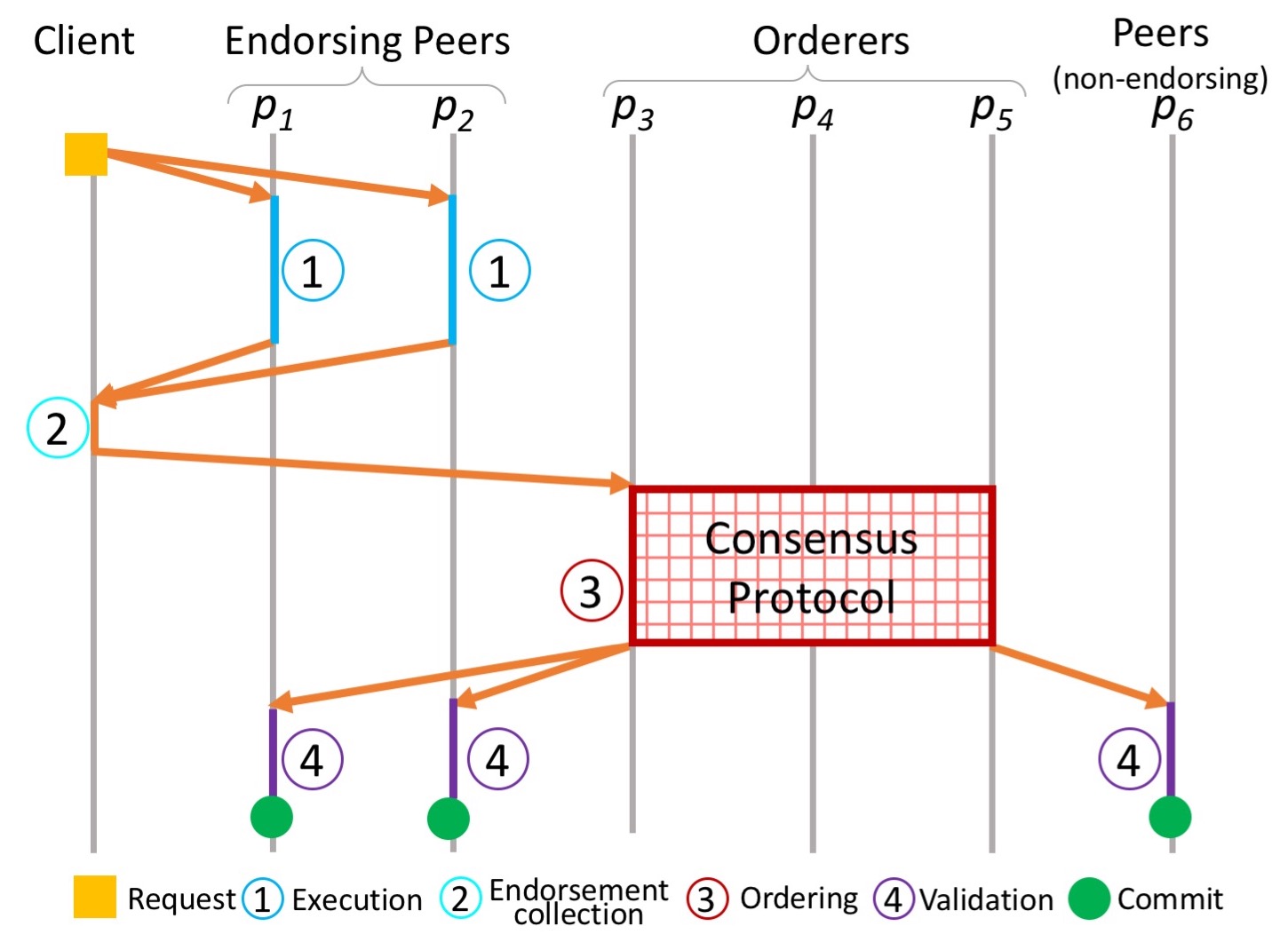}}
  \hfill
  \caption{Existing Paradigms for Blockchains}
  \label{fig:architecture}
  \vspace{-1.5em}
  \end{figure*}

Existing blockchain systems can be divided into two main categories:
permissionless blockchain systems, e.g., Ethereum (with PoS-based consensus) \cite{ethereum17} and
permissioned blockchain systems, e.g., Tendermint  (with BFT-type consensus) \cite{kwon2014tendermint}.

Permissionless blockchains are public, and anyone can participate
without a specific identity.
Permissionless blockchains mainly follow the order-execute paradigm where
nodes validate the transactions,
put the transactions into blocks,
and try to solve some cryptographic puzzle.
The lucky peer that solves the puzzle
multicasts the block to all nodes.
When a node receives a block of transactions,
it validates the solution to the puzzle and all transactions in the block.
Then, the node executes the transactions within a block sequentially.
Such a paradigm requires all nodes to execute every transaction and
all transactions to be deterministic.

Figure~\ref{fig:oe} shows the transaction flow
for a permissionless blockchain.
When a peer receives transactions from clients,
in step $1$, the peer validates the transactions, puts them into a block, and
tries to solve the cryptographic puzzle.
If the peer is lucky ($p_3$ in the figure) and solves the puzzle before other peers, it multicasts the block to all the peers.
All the nodes then validate the block and its transactions (step $3$),
execute the transactions sequentially (step $4$), and finally, update their respective copies of the ledger.
Note that if multiple peers solve the puzzle at the same time, a fork happens in the blockchain.
However, once a block is added to either of the fork branches, nodes in the network join the longest chain.

A permissioned blockchain, on the other hand, consists of a set of known, identified nodes but which do not fully trust each other.
In permissioned blockchains, since the nodes are known and identified,
traditional consensus protocols can be used to order the requests \cite{cachin2016architecture}.


A permissioned blockchain can follow either order-execute or execute-order paradigm.
In order-execute permissioned blockchains, as can be seen in Figure~\ref{fig:oe-perm},
a set of peers (might be all of them) validate the transactions,
agree on a total order for the transactions,
put them into blocks and multicast them to all the nodes.
Each node then validates the block, executes the transactions using a "smart contract", and updates the ledger.
A {\em smart contract} is 
a computer program that self-executes once it is established and deployed.
Smart contracts are similar to \textit{databases triggers} where the logic of 
the contract is triggered to be executed once some conditions or terms are met.
They have the advantages of supporting real-time updates, accurate execution, and little human intervention.

In order-execute permissioned blockchains, similar to order-execute permissionless blockchains,
every smart contract runs on every node.
Smart contracts include the logic of applications and it might be desirable to
restrict access to such contracts.
While cryptographic techniques are used to achieve confidentiality,
the considerable overhead of such techniques makes them impractical \cite{androulaki2018hyperledger}.
Furthermore the sequential execution of transactions on every node
reduces the blockchain performance in terms of throughput and latency.

In contrast to the order-execute paradigm,
Hyperledger Fabric \cite{androulaki2018hyperledger} presents a new paradigm
for permissioned blockchains by switching the order of execution and ordering.
The execute-order paradigm was first presented in Eve \cite{kapritsos2012all}
in the context of Byzantine fault-tolerant SMR.
In Eve peers execute transactions concurrently and then verify
that they all reach the same output state, using a consensus protocol.
In fact, Eve uses an Optimistic Concurrency Control (OCC) \cite{kung1981optimistic}
by assuming low data contention where conflicts are rare.

Hyperledger Fabric uses a similar strategy;
a client sends a request to a subset of peers, called endorsers
(the nodes that have access to the smart contract).
Each endorser executes the request and sends the result back to the client.
When the client receives enough endorsements (specified by some endorsement policy),
it assembles a transaction including all the endorsements and
sends it to some specified (ordering) peers to establish a {\em total order} on all transactions.
This set of nodes establishes consensus on transactions, creates blocks, and
broadcasts them to every node.
Finally, each peer validates a transaction within a received block
by checking the endorsement policy and read-write conflicts and then updates the ledger.
Since a validation phase occurs at the end, the paradigm is
called {\em execute-order-validate}.
Figure~\ref{fig:eo} presents the flow of transactions in Fabric.
Note that in Fabric the consensus protocol is pluggable and the system
can use a crash fault-tolerant protocol, e.g., Paxos \cite{lamport2001paxos},
a Byzantine fault-tolerant protocol, e.g., PBFT \cite{castro2002practical},
or any other protocol.

While Fabric solves the confidentiality issue by
executing each transaction on a specified subset of peers (endorsers) and
increases the performance of blockchains
by executing the transactions in parallel
(instead of sequentially as the order-execute paradigm does),
it performs poorly on workloads
with high-contention, i.e.,
many {\em conflicting transactions} in a block,
due to its high abort rate.

Two transactions {\em conflict} if they access the same data and
one of them is a write operation.
In such a situation,
the order of executing the transactions is important,
indeed, the later transaction in a block has to wait for
the earlier transaction to be executed first.
As a result, if two conflicting transactions execute in parallel, the result is
invalid.
Although Fabric guarantees correctness by checking the conflicts 
in the validation phase (the last phase) and
disregarding the effects of invalid transactions,
the performance of the blockchain is highly reduced by such conflicts.

%% file: sec_model.tex
\section{The \orexp Paradigm}\label{sec:model}

In this section, we introduce \orexp, a new order-execute paradigm for permissioned blockchains.
\orexp is mainly designed to support distributed applications with high-contention workloads.

\orexp consists of a set of nodes in an asynchronous distributed network where
each node has one of the following roles:

\begin{itemize}

\item {\em Clients} send operations to be executed by the blockchain.

\item {\em \Ords} agree on a total order of all transactions.

\item {\em \Exes} validate and execute transactions.

\end{itemize}
    
The set of nodes in \orexp is denoted by $N$ where
$O$ of them are \ords, and
$E$ of them are \exes.

\orexp supports distributed applications running concurrently on the blockchain.
For each application a program code including the logic of that application ({\em smart contract})
is installed
on a (non-empty) subset of \exe peers called the {\em agents} of the application.
We use ${\cal A}=\{A_1, ...,A_n\}$ to denote the set of applications (ids) and
$\Sigma(A_i)$ to specify the non-empty set of agents of each application $A_i$ where
$\Sigma:{\cal A} \mapsto 2^E - \emptyset$.
Every peer in the blockchain knows the agents of each application and
the set of \ords.

Each pair of peers is connected with point-to-point
bi-directional communication channels.
Network links are pairwise authenticated, which guarantees
that a Byzantine node cannot forge a message from a correct node,
i.e., if node $i$ receives a message $m$ in the incoming link from node $j$,
then node $j$ must have sent message $m$ to $i$ beforehand.

\subsection{\Ords} \label{sec:ord}

Checking accesses, ordering the requests, constructing blocks, generating dependency graphs, and multicasting the blocks
are the main services of \ords in the \orexp paradigm.

Since multiple applications run on the blockchain and each application might have its own set of clients,
\ords act as trusted entities to restrict the processing of requests that are sent by unauthorized clients.
If a client is not authorized to perform an operation on the requested application, \ords simply
discard that request. \Ords also check the signature of the requests to ensure their validity.

\Ords use an asynchronous fault-tolerant protocol to establish consensus.
Fault-tolerant protocols use the state machine replication algorithm \cite{lamport1978time}
where replicas agree on an ordering of incoming requests.
The algorithm has to satisfy two main properties,
(1) {\em safety}: all correct nodes receive the same requests in the same order, and
(2) {\em liveness}: all correct client requests are eventually ordered.
Fischer et al. \cite{fischer1985impossibility} show that 
in an asynchronous system, where nodes can fail,
consensus has no solution that is both safe and live.
Based on that impossibility result, most fault-tolerant protocols
satisfy safety without any synchrony assumption and
consider a synchrony assumption to satisfy liveness.

\orexp, similar to Fabric \cite{androulaki2018hyperledger},
uses a pluggable consensus protocol for ordering, thus resulting in a modular paradigm.
Depending on the characteristics of the network and peers 
\orexp might employ a Byzantine, a crash, or a hybrid fault-tolerant protocol.
The number of \ords is also determined by the utilized protocol and
the maximum number of simultaneous failures in the network.
For example, crash fault-tolerant protocols, e.g., Paxos \cite{lamport2001paxos},
guarantee safety (consistency) in an asynchronous network using $2f{+}1$ nodes
to overcome the simultaneous crash failure of any $f$ nodes while
in Byzantine fault-tolerant protocols, e.g., PBFT \cite{castro2002practical},
$3f{+}1$ nodes are needed to provide safety in the presence of $f$ malicious nodes.
Furthermore,
\ords do not have access to any smart contract or the application state,
nor do they participate in the execution of transactions.
This makes \ords independent of the other peers and adaptable to a changing environment.

\Ords batch multiple transactions into blocks.
Batching transactions into blocks improves the performance of the blockchain
by making data transfers more efficient especially in a geo-distributed setting.
It also amortizes the cost of cryptography.
The batching process is {\em deterministic} and therefore produces
the same blocks at all \ords.

\begin{figure}[t] \center
\includegraphics[width=\linewidth]{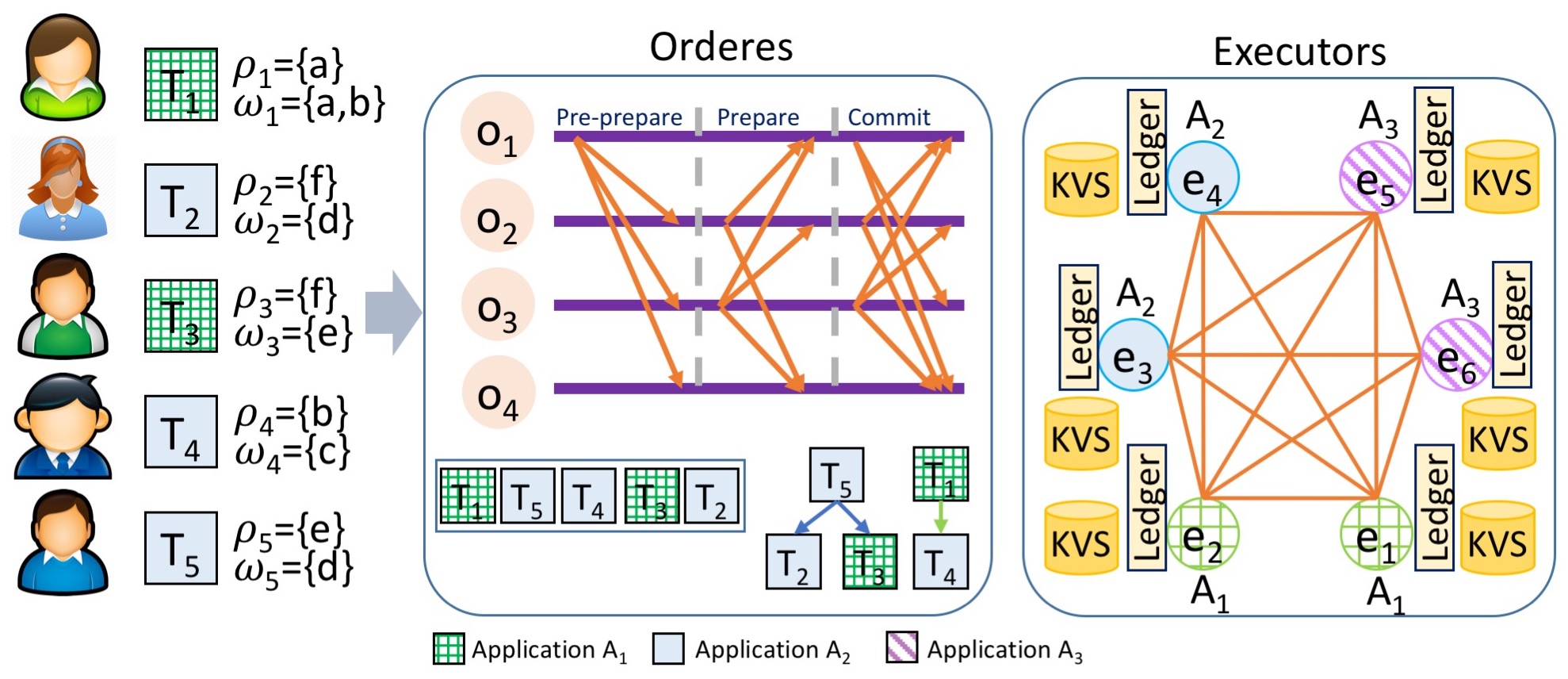}
\caption{The Components of \orexp Paradigm}
\label{fig:arch}
\vspace{-1.5em}
\end{figure}

Figure~\ref{fig:arch} shows the components of the \orexp paradigm.
As can be seen, clients send requests (transactions) to be executed by different applications.
Here, transactions $T_1$ and $T_3$ are for some application $A_1$ and $T_2$, $T_4$, and $T_5$ are for another application $A_2$.
The \ords, i.e., $o_1$, $o_2$, $o_3$, and $o_4$, then order the transactions
and put them into a block. In the figure, \ords use PBFT \cite{castro2002practical} to order the requests.
The resulting block contains five transactions which are ordered as $[T_1, T_5, T_4, T_3, T_2]$.

Next, \ords generate a "dependency graph" for the transactions within a block.
In order to generate dependency graphs a priori knowledge of transactions' read- and write-set is needed.
Each transaction consists of a sequence of
reads and writes, each accessing a single record.
Here we assume that the
read-set and write-set are pre-declared or can be obtained from the transactions via a static analysis, e.g.,
all records involved in a transaction are accessed by their primary keys.
Note that even if that assumption does not hold, the system can employ other techniques
like speculative execution \cite{faleiro2017high}
to obtain the read-set and write-set of each transaction.

Given a transaction $T$,
$\omega(T)$ and $\rho(T)$
are used to represent the set
of records written and read, respectively.
Each transaction $T$ is also associated with a timestamp $ts(T)$
where for each two transactions $T_i$ and $T_j$ within a block
such that $T_i$ appears before $T_j$,
$ts(T_i) < ts(T_j)$.

We define "ordering dependencies" to show possible conflicts
between two transactions from the same or different applications.
Two transactions conflict if they access the same data and
one of them is a write operation.

\begin{deff}
Given two transactions $T_i$ and $T_j$.
An {\em ordering dependency} $T_i \succ T_j$ exists if and only if
$ts(j) > ts(i)$ and one of the following hold:

\begin{itemize}
\item $\rho(T_i) \cap \omega(T_j) \ne \varnothing$
\item $\omega(T_i) \cap \rho(T_j) \ne \varnothing$
\item $\omega(T_i) \cap \omega(T_j) \ne \varnothing$
\end{itemize}

\end{deff}

\begin{deff}
Given a block of transactions, the {\em dependency graph} of the block is a directed graph $G=(\caT,\caE)$ where
$\caT$ is the set of transactions within the block and
$\caE= \{(T_i,T_j) \mid T_i \succ T_j\}$
\end{deff}

We use the example in Figure~\ref{fig:arch} to illustrate the dependency graph construction process.
As can be seen the block consists of  five transactions
which are ordered as $[T_1, T_5, T_4, T_3, T_2]$, i.e.,
$ts(T_1) < ts(T_5) < ts(T_4) < ts(T_3) < ts(T_2)$.
Since  $T_4$ reads data item $b$ which is written by an earlier transaction $T_1$ there is an
ordering dependency $T_1 \succ T_4$, thus $(T_1,T_4)$ is an edge of the dependency graph.
Similarly, $T_2$ writes data item $d$ which is also written by $T_5$ ($T_5 \succ T_2$) and
$T_3$ writes data item $e$ which is read by $T_5$
($T_5 \succ T_3$).
As a result,
edges $(T_5,T_2)$ and $(T_5,T_3)$ are also in the graph.

The constructed graph can be used by the \exes to manage the execution of transactions.
In particular, transactions that are not connected to each
other in the dependency graph, e.g., $T_1$ and $T_2$, can be processed concurrently by independent execution threads.

The dependency graph generator is an independent module
in the \orexp paradigm. Therefore, it can also be adapted to 
a multi-version database system \cite{bernstein1983multiversion}.
In a multi-version database, each write creates a new version
of a data item, and reads are directed
to the correct version based on the
position of the corresponding transaction in the block (log).
Since writes do not overwrite each other, the system has more flexibility to manage the order of reads and writes.
As a result,
for any two transactions $T_i$ and $T_j$ within a block
where $T_i$ appears before $T_j$,
$T_i$ and $T_j$ can concurrently write the same data item or
$T_i$ reads and $T_j$ writes the same data item.
However, if $T_i$ wants to write and $T_j$ wants to read
the same data item, they cannot be executed in parallel.

It should be noted that in some dependency graph construction approaches, e.g., DGCC \cite{yao2015dgcc}, 
transactions are broken down into
transaction components, which allows the system to parallelize the
execution at the level of operations.
The dependency graph generator module in \orexp can
also be designed in a similar manner.

A dependency graph has two main benefits.
First, it reduces the abort rate in comparison to the execute-order paradigm
by exposing conflicts between transactions.
Recall that
conflicting transactions reduce the performance of the execute-order paradigm,
because if two conflicting transactions are executed simultaneously,
the system has to abort the later one.
A dependency graph gives a partial order based on the conflicts between
transactions and as long as the transactions
are executed in an order consistent with the dependency graph, the results are valid.
Second, dependency graphs enable higher concurrency.
Since a dependency graph provides a partial order between transactions within a block,
non-conflicting transactions can still be executed in parallel.
Such parallelism improves the performance of \orexp paradigm in comparison to
the traditional order-execute paradigm where transactions are executed sequentially.

When the dependency graph is generated, \ords multicast a message
including the block and its dependency graph to all \exes.
Depending on the employed consensus protocol, either the leader or all the \ords multicast the message.

\subsection{\Exes} \label{sec:exe}

Executing and validating transactions,
updating the ledger and the blockchain state, and multicasting the blockchain state after executing transactions
are the main services of \exe peers.
\Exes in OXII correspond to the endorsers in Hyperledger \cite{androulaki2018hyperledger}.
Each \exe peer maintains three main components:
(1) The blockchain ledger,
(2) The blockchain state, and
(3) Some smart contracts.

The blockchain ledger is
an append-only data structure recording all transactions in the form of a hash chain.
When a block of transactions is executed and validated, each \exe peer
appends the block to its copy of the ledger.

Each \exe node is an agent for one or more applications where for each application
the smart contract of that application, i.e.,
a program code that implements the application logic,
is installed on the node.
When an \exe receives a block from the \ords, it checks the application of the transactions within the block.
If the \exe is an agent for any of the transactions, it executes the transactions on the corresponding smart contract
following the dependency graph.
In fact, the \exe confirms the order of dependent transactions and executes independent transactions in parallel.
Finally, it multicasts the execution results (updated blockchain state) to all other peers.

For each transaction within a block where the \exe is not an agent of the transaction,
the \exe waits for matching updates from a specified number of \exes,
who are the agents of the transaction, before committing the update.
These numbers are decided by the system and known to all \exes
(similar to endorsement policies in Hyperledger).
We use $\tau(A)$ to denote the required number of matching updates
for the transactions of application $A$.

In Figure~\ref{fig:arch}, \exe nodes $e_1$ and $e_2$ are the agents of application $A_1$
(with transactions $T_1$ and $T_3$) and
\exe nodes $e_3$ and $e_4$ are the agents of application $A_2$
(with transactions $T_2$, $T_4$ and $T_5$).


%% file: sec_algorithm.tex
 \section{\system}\label{sec:system}

In this section, we present \system, a permissioned blockchain 
designed specifically in the \orexp paradigm.
We first give a summary of \system and then explain the ordering and execution phases in detail.

\subsection{\system Overview}

\begin{figure}[t] \center
\includegraphics[width=0.8\linewidth]{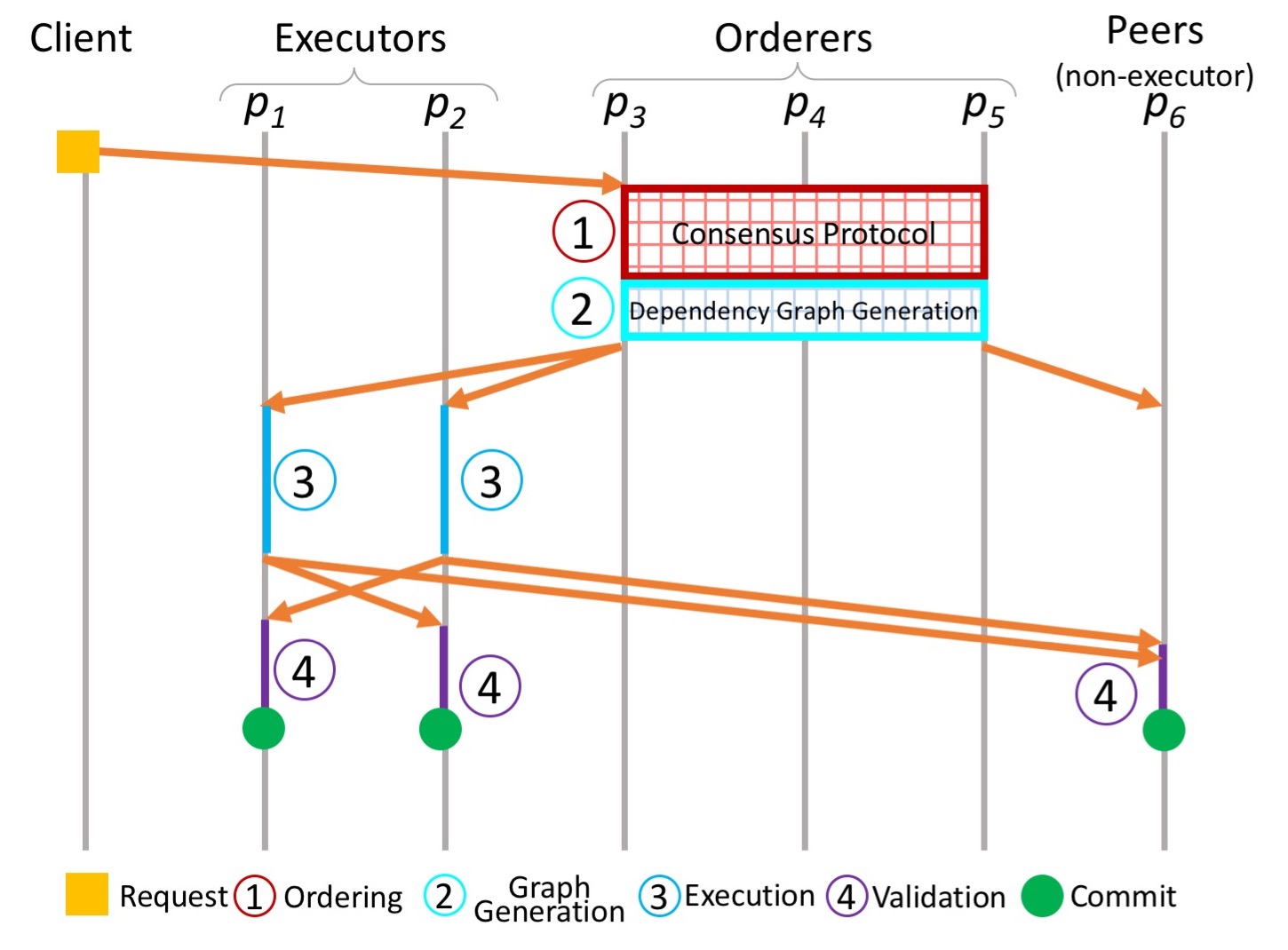}
\caption{The Flow of Transactions in \system}
\label{fig:oxii}
\vspace{-1.5em}
\end{figure}

\system is a permissioned blockchain
to execute distributed applications.
\system is designed in the \orexp paradigm.
In particular, the normal case operation for \system
to execute client requests proceeds as follows.

Clients send requests to the \ord nodes and 
the \ords run a consensus algorithm among themselves to reach agreement on the order of transactions.
\Ords then construct a block of transactions and
generate a dependency graph for the transactions within the block.

Once the dependency graph is generated, the block along with the graph is multicast
to all the \exe nodes.
The \exes which are the agents of some transactions within the block, execute the corresponding transactions and multicast the results, i.e., updated records in the datastore, to every \exe node.
Each \exe node in the network waits for the required number of matching results from the \exes
to update the ledger and blockchain state (datastore).
The required number of matching results for each application
is determined by the system and might be different for different applications.

The flow of transactions in \system can be seen in Figure~\ref{fig:oxii} where $p_3$, $p_4$, and $p_5$ are the \ord nodes, and $p_1$, $p_2$, and $p_6$ are the \exe nodes from which
$p_1$ and $p_2$ are the agents for the requests.
Upon receiving requests from clients, \ords order the requests, put them into a block, generate the dependency graph for the block, and multicast the block along with the graph to all the \exe nodes. The agents of the corresponding application ($p_1$ and $p_2$) execute the transactions and multicast the updated state of the blockchain to the other \exe nodes. Upon receiving the required number of matching messages for each transaction, each \exe commits (or aborts) the transaction by updating the blockchain state.
The block is also appended to the ledger.

\subsection{Ordering Phase}

The goal of the ordering phase is to establish a total order
on all the submitted transactions.
A client $c$ requests an operation $op$ for some application $A$
by sending a message
$\big\langle\text{\REQ}, op, A, ts_c, c\big\rangle_{\sigma_c}$
to the \ord $p$ it believes to be the {\em primary}
(an \ord node that initiates the consensus algorithm).
Here, $ts_c$ is  the client's timestamp and the entire message
is signed with signature $\sigma_c$.
We use timestamps of clients to totally order the requests of each client and to ensure exactly-once semantics for the execution of client requests.

Upon receiving a client request, the primary \ord $p$ checks the signature to ensure it is valid,
makes sure the client is allowed to send requests for application $A$ (access control), and then
initiates a consensus algorithm
by multicasting the request to other \ords.
Depending on the utilized consensus protocol, several rounds of communication occurs between \ords to
establish a total order on transactions.

Once the \ords agree on the order of a transaction, they put
the transaction in a block.
Batching multiple transactions into blocks
improves the throughput of the broadcast protocol.
Blocks have a pre-defined maximal size, maximal number of transactions, and
maximal time the block production takes since the first transaction of a new block was received.
When any of these three conditions is satisfied, a block is full.
Since transactions are received in order, the first two conditions are deterministic.
In the third case, to ensure that the produced blocks by all \ords are the same, 
the primary sends a {\em cut-block} message in the consensus step of the last request.

When a block is produced, \ords generate a dependency graph for the block as explained in Section~\ref{sec:ord}.
Generating dependency graphs requires 
a priori knowledge of transactions' read- and write-set.
Here, we assume that the requested operations include the read- and write-set.

When the graph is constructed, each \ord node $o$ multicasts a message
$\big\langle\text{\NEWB},n, B, G(B), {\bf A}, o, h\big\rangle_{\sigma_o}$
to all \exe nodes where
$n$ is the sequence number of the block,
$B$ is the block consisting of the \req messages,
$G(B)$ is the dependency graph of $B$,
${\bf A}$ is the set of applications that have transactions in the block, and
$h = H(B')$ where $H(.)$ denotes the cryptographic hash function and $B'$ is
the block with sequence number $n{-}1$.

\subsection{Execution Phase}\label{sec:execution}

Each request for an application is executed on the specified set of \exes, i.e., agents of that application.
Upon receiving a \newb message
$\big\langle\text{\NEWB}, n, B, G(B),\allowbreak {\bf A}, o, h\big\rangle_{\sigma_o}$
from some \ord $o$,
\exe $e$ checks the signature and the hash to be valid and logs the message.
It also checks the set {\bf A} to see if the block contains any transaction that
needs to be executed by the node, i.e.,
an application $A_i \in {\bf A}$ such that $e \in \Sigma(A_i)$.

\begin{algorithm}[t]
\footnotesize
\caption{Execution of Transactions on an \exe $e$}
\label{alg:exe}
\textbf{Input:} A block $B$ and its dependency graph $G(B)$
\begin{algorithmic}[1]
\State Initiate Set $W_e$ to be empty 
\For{transaction $x$ in $B$}
    \If{$e$ is an agent of $x$'s application}
        \State Add $x$ to $W_e$
    \EndIf
\EndFor
\While {$W_e$ in not empty}
    \For{transaction(node) $x$ in $W_e$}
        \If{all $Pre(x)$ are in $C_e \cup X_e$}
            \State trigger Execute($x$)
        \EndIf
    \EndFor
\EndWhile
\end{algorithmic}
\end{algorithm}

When an \exe node receives a specified number of matching \newb messages, e.g., $f+1$ messages if the consensus protocol is PBFT,
it marks the \newb as a valid block and enters the execution phase.
The execution phase consists of three procedures that are run concurrently:
(1) Executing the transaction following the dependency graph,
(2) Multicasting \cmt messages including
the execution results to other \exe nodes, and
(3) Updating the blockchain state upon receiving \cmt messages from a sufficient number of \exe nodes.

If an \exe node is not an agent of any transaction within the block,
the node becomes a {\em passive} node and only the third procedure is run to update the blockchain state.
However, if a node is an agent for some transaction's application in the block,
it runs all three procedures; executes the corresponding transactions following the dependency graph,
multicasts the results, and also updates the blockchain state.

A transaction can be executed only if all of its "predecessors" in
the dependency graph are committed.
We define functions $Pre$ and $Suc$ to present the
set of predecessors and successors of a node in a dependency graph
respectively.
More formally, 
Given a dependency graph $G=(\caT,\caE)$, and 
a node (transaction) $x$ in $\caT$,
$Pre(x)=\{y \mid (y,x) \in \caE\}$ and
$Suc(x)=\{y \mid (x,y) \in \caE\}$.

The execution procedure on a node $e$ is shows in 
Algorithm~\ref{alg:exe}.
An empty set $W_e$ is initiated to keep all the transactions
that will be executed by \exe $e$, i.e.,
$e$ is an agent for the application of those transactions.
Set $X_e$ stores the executed transactions by $e$ and
$C_e$ keeps the committed transactions.
For each transaction $x$ in $W_e$, the procedure checks
the predecessors of $x$, if $x$ has no predecessor, or all of its predecessors are executed by $e$ or committed,
transaction $x$ is ready to be executed, so 
an execution thread is triggered.

\begin{algorithm}[t]
\footnotesize
\caption{Multicasting the Results}
\label{alg:multicast}

\begin{algorithmic}[1]
\State Initialize set $X_e$ to be empty
\State $cut$ = false
\State Upon obtaining an execution result $(x,r)$
\State \quad Add pair $(x,r)$ to $X_e$
\State \quad Remove $x$ from $W_e$
\For{$y$ where $(x,y)$ is an edge in $G(B)$}
        \If {$y$'s application is different from  $x$'s application}
            \State $cut$ = true
            \State break
        \EndIf
\EndFor
\If {$cut$ = true}
    \State Multicast
    $\big\langle\text{\CMT},X_e,e\big\rangle_{\sigma_e}$
    to all \exes
    \State Clear $X_e$
\EndIf
\end{algorithmic}
\end{algorithm}

To multicast the execution results depending on the transactions' applications three different situations could happen.
If all the transactions within a block belong to the same application, an agent $e$ executes all of the transactions
following the dependency graph and
multicast a {\em \cmt} message
$\big\langle\text{\CMT},S,e\big\rangle_{\sigma_e}$
to all other \exe nodes.
Here, $S$ presents the state of the blockchain and
consists of a set of pairs $(x,r)$ where
$x$ is a transaction (id) and $r$ is the set of updated records resulting from the execution of $x$
on the datastore.
Note that if a transaction $x$ is not valid, the \exe puts ($x$,{\tt "abort"}) in $S$.

If the transactions within a block are for different applications but the transactions of each application
access a disjoint set of records,
the agents still can execute the corresponding transactions independently and multicast
a single \cmt message with all the results to other \exe nodes.
In this case, the dependency graph is disconnected and can be decomposed to different components where
the transactions of each component are for the same application and there is no edge that connects any two components.

However, if there are some dependencies between the transactions of two applications, the agents of those two applications cannot execute the transactions independently.
In fact, the agents of one application have to wait for the agents of other applications to execute all their transactions and send the \cmt message.
In this case, a deadlock might occur.

\begin{figure}[t] \center
\includegraphics[width=0.9\linewidth]{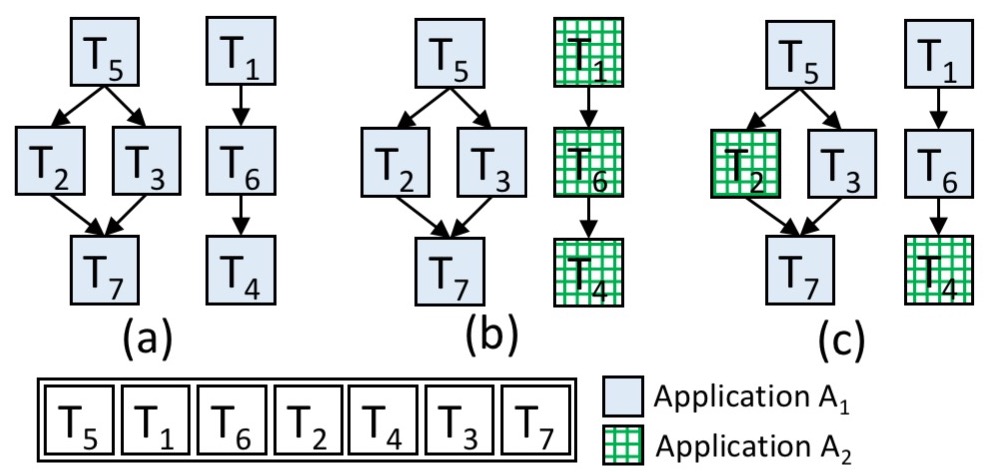}
\caption{Three Dependency Graphs}
\label{fig:graph}
\vspace{-1.5em}
\end{figure}

Figure~\ref{fig:graph} shows three dependency graphs for
a block of seven transactions $T_1$ to $T_7$.
In Figure~\ref{fig:graph}(a), all the seven transactions belong to the same application, $A_1$.
Therefore, the agents of application $A_1$ can execute the transactions following the dependency graph and multicast the results of all transactions together when they all are executed.
In Figure~\ref{fig:graph}(b) although the transactions belong to different applications
($T_2$, $T_3$, $T_5$, and $T_7$ are for application $A_1$ and
$T_1$, $T_4$ and $T_6$ are for application $A_2$),
there is no dependency between the transactions of application $A_1$ and the transactions of application $A_2$.
As a result, the agents can still execute independently and multicast the results once the execution of their transactions is completed.
However, in Figure~\ref{fig:graph}(c) since there are some dependencies between the transactions of the two applications, the agents cannot execute their transactions independently.
For example, to execute transaction $T_2$, the agents of application $A_2$ need the execution results of transaction $T_5$ from the agents of $A_1$.
Similarly, transaction $T_4$ cannot be executed before committing the execution results of transaction $T_6$.

To prevent a deadlock situation, one possibility is that agents send a \cmt message as soon as the execution of each transaction is completed.
While this approach solves the blocking problem, the number of
exchanged \cmt messages will be large. Indeed, if a block includes $n$ transactions and
each application has on average $m$ agents, there will be total $n*m$ exchanged \cmt messages for the block.

A more efficient way is to send \cmt messages when the execution results are needed by some other agents.
Basically, an agent keeps executing the transactions and collecting the results until
the results of an executed transaction is needed by some other transactions which belong to other applications.
At that time, the agent generates a \cmt message including the results
of all the executed transactions and multicasts it to all \exe nodes.
Upon receiving a \cmt message from an \exe, the node
validates the signature and logs the message.
Once the node receives the specified number of matching results
for a transaction, the results are reflected in the datastore and
the transaction is marked as committed.

\begin{algorithm}[t]
\footnotesize
\caption{Updating the Blockchain State}
\label{alg:update}

\begin{algorithmic}[1]
\For{transaction $x$ in $B$}
    \State Initialize set $R_e(x)$ to be empty
\EndFor
\State Initialize set $C_e$ to be empty
\State Upon Receiving a valid
$\big\langle\text{\CMT},S,n\big\rangle_{\sigma_n}$ message
    \For{valid $(x,r) \in S$}
        \State Add $(r,n)$ to $R_e(x)$
        \If {Matching records in $R_e(x) \geq \tau(A)$}
            \State Update the blockchain state
            \State Add $x$ to $C_e$
        \EndIf
    \EndFor

\end{algorithmic}
\end{algorithm}

Algorithm~\ref{alg:multicast} presents the multicasting procedure on a node $e$.
An empty set $X_e$ is initiated to store the results of the executed transactions.
When the execution of a transaction $x$ is completed,
the execution result $(x,r)$ is added to $X_e$ and 
transaction $x$ is removed from the waiting transactions $W_e$.

Then, the procedure checks all the successor nodes of $x$ in the dependency graph.
If any of the successor nodes of $x$ belongs to an application
different from the application of $x$,
the execution result of transaction $x$ might be needed by other agents, thus a multicasting has to occur.
To do so, node $e$ removes all the stored results from $X_e$
and puts them in a \cmt message and multicast the \cmt message to all other \exe nodes.

For example, in Figure~\ref{fig:graph}(c), upon executing
the transaction $T_5$, since $T_5$ has a successor node $T_2$ that belongs to another application,
the \exe node multicasts a \cmt message including the execution results of $T_5$ to all other \exe nodes.
Note that if $T_1$ is already executed, the \exe node
puts the execution results of $T_1$ in the \cmt message as well.
Similarly, when the execution of $T_6$ is completed,
the \exe node multicasts a \cmt message including the execution results of $T_6$ and any other
executed but not yet multicast transactions.

Finally, the updating procedure receives \cmt messages from other
\exe nodes and updates the blockchain state.
The updating procedure on a node $e$ is presented in 
Algorithm~\ref{alg:update}.
the procedure first initializes an empty set $R_e(x)$
for each transaction $x$ in the block.
It also initializes an empty set $C_e$ to collect committed transactions.
When node $e$ receives a commit message
$\big\langle\text{\CMT},S,n\big\rangle_{\sigma_n}$ from some \exe $n$,
it checks the signature to be valid and then checks the
set $S$.
Recall that $S$ consists of pairs of transactions and their execution results.
For each pair $(x,r)$, it first checks whether
node $n$ is an agent for the application of transaction $x$ and then
a pair of $(r,n)$, i.e., execution results and the \exe
to $R_e(x)$.
Assuming $A$ is the $x$'s application,
If the number of the matching tuples in $R_e(x)$
is equal to $\tau(A)$, i.e., the specified number of messages
for the transaction's application,
the execution results are valid and can be committed.
As a result, the procedure updates the blockchain state (datastore)
and adds the transaction $x$ to the committed transactions $C_e$.

%% file: sec_exp.tex
\section{Experimental Evaluations}\label{sec:eval}

In this section, we conduct several experiments to evaluate different paradigms for permissioned blockchains.
We discussed the two existing paradigms for permissioned blockchains in Section~\ref{sec:back}:
sequential order-execute (OX) where requests are ordered and then executed sequentially on every node, and
execute-order-validate (XOV) introduced by Hyperledger Fabric \cite{androulaki2018hyperledger}
where requests are executed by the agents of each application, ordered by the ordering service,
and validated by every peer.
We implemented two permissioned blockchain systems specifically designed in the OX and XOV paradigms
as well as \system that is designed in the \orexp paradigm.
It should be noted that our implementation of XOV is different from the Hyperledger fabric system.
Hyperledger is a distributed operating system and includes many components which are not the focus of our evaluations.
In fact, the purpose of our experiments is to compare the architectural aspect of the blockchain systems,
thus, all three systems are implemented using the same programming language ({\em Java}).
To have a fair comparison, we also used similar libraries and optimization techniques for all three systems as far as possible.

We implemented a simple accounting application where each client has several accounts.
Each account can be seen as a pair of $(amount, PK)$ where $PK$ is the public key of the owner of the account.
Clients can send requests to transfer assets from one or more of their accounts to other accounts.
For example, a simple transaction $T$ initiated by client $c$ might
"transfer $x$ units from account $1001$ to account $1002$".
The transaction is valid if $c$ is the owner of account $1001$
and the account balance is at least $x$.
Here the read-set of transaction $T$ is $\rho(T)=\{1001\}$ and its write-set is $\omega(T)=\{1001,1002\}$.
A transaction might read and write several records.

\input{graph_block}

The experiments were conducted on the Amazon EC2 platform.
Each VM is Compute Optimized c4.2xlarge instance with 8 vCPUs and 15GM RAM,
Intel Xeon E5-2666 v3 processor clocked at 3.50 GHz.
For \ords, similar to Hyperledger \cite{androulaki2018hyperledger},
we use a typical Kafka orderer setup with 3 ZooKeeper nodes, 4
Kafka brokers and $3$ \ords, all on distinct VMs.
Unless explicitly mentioned differently,
there are three applications in total each with a separate \exe (endorser) node.

\input{graph_contention}

When reporting throughput measurements, we use an increasing
number of clients running on a single VM,
until the end-to-end throughput is saturated,
and state the throughput just below saturation. Throughput
numbers are reported as the average measured during the steady
state of an experiment.

\subsection{Choosing the Block Size}
An important parameter that impacts both throughput and latency is the block size.
To evaluate the impact of the block size on performance,
in this set of experiments, assuming that the transactions have the same size,
we increase the number of transactions in each block from $10$ to $1000$.
For each block size, the peak throughput and the corresponding average end-to-end latency is measured.
As can be seen in Figure~\ref{fig:block}, by increasing the number of transactions per block till ${\sim} 200$,
the throughput of \orexp increases, however, any further increasing reduces the throughput due to the
large number of required computations for the dependency graph generation.
Similarly, by increasing the number of transactions per block till ${\sim} 200$, the delay decreases.
Afterward, adding more transactions to the dependency graph becomes more time consuming than multicasting the block.
As a result, \orexp is able to process more than $6000$ transactions in $78 ms$ with $200$ transactions per block.
In the OX paradigm, since nodes execute transactions sequentially, the block creation time
is negligible in comparison to the execution time, thus other than in the first experiment,
increasing the number of transactions per block
does not significantly affect the throughput and latency.
In the XOV paradigm, since \exes (endorsers) of the three applications can execute the transactions in parallel,
the performance is better than OX (twice as much as OX in its peak throughput).
However, its performance is still much less than \orexp, i.e.,
the peak throughput of XOV is $30\%$ of the peak throughput of \orexp as
\orexp can execute many (and not only three) non-conflicting transactions in parallel.
As can be seen, the peak throughput of XOV is obtained in ${\sim} 100$ transactions per block.
\vspace{-1em}

\subsection{Performance in Workloads with Contention}
In the next set of experiments, we measure the performance of all three paradigms for
workloads with different degrees of contention. we consider
no-contention,
low-contention ($20\%$ conflict),
high-contention ($80\%$ conflict), and
full-contention workloads
where the results are shown in Figure~\ref{fig:contention}(a)-(d) respectively.
Note that the dependency graph of each block in the first workload has no edge whereas
the dependency graph of each block in the last workload is a chain.
In OX and OXII there are $200$ transactions per block and
for XOV, we keep changing the block size to find its peak throughput.
Contentions could happen between the transactions of the same application
or the transactions of different applications (if they access shared data).
In OX, since nodes execute transactions sequentially, there is no difference between these two types of contention.
In XOV also, since the execution is the first phase,
there is no much difference between contention within an application or across
applications and they both result in transaction abort.
In OXII, however, as discussed in Section~\ref{sec:execution}, for  contention across applications,
the agents of different applications communicate
to each other during the execution of a block of transactions, thus the performance is affected.
As a result, in this set of experiments, 
for each workload, we report the performance of OX, XOV, OXII with conflicting transactions within an application, and
OXII with conflicting transactions across applications (the dashed line).

As mentioned earlier, in the OX paradigm, transactions are executed sequentially. As a result,
the performance of OX remains unchanged in different workloads.
XOV can execute $3$ (number of applications) transactions in parallel and since the workload has no-contention,
the execution results are valid.
OXII, on the other hand, significantly benefits from no-contention workloads by executing the transactions in parallel.
As shown in Figure~\ref{fig:contention}(a), OXII executes more than $6000$ transactions with latency less than $80$ ms
whereas the peak throughput of OX is $900$ transactions with more than $500$ ms latency.
XOV can also execute around $1800$ transactions in $600$ ms
($70\%$ less throughput and $7.5$ times latency in comparison to OXII).
Since the workload has no conflicting transaction, there is no contention across applications.

By increasing the degree of contention
(Figure~\ref{fig:contention}(b) and Figure~\ref{fig:contention}(c)),
the throughput of XOV decreases dramatically, e.g.,
the peak throughput of XOV in a high-contention workload is around $25\%$ of its peak throughput in a no-contention workload.
This decrease is expected because XOV validates and aborts the conflicting transactions at the very end (last phase).
The throughput of OXII is also affected by increasing the degree of contention, however, it still shows better performance
than both OX and XOV, i.e., OXII is still able to process $1600$ transactions in sub-second latency whereas
OX and XOV process $900$ and $350$ transactions respectively.
Processing the workloads with contention across the applications decreases the performance of OXII
due to the increasing rounds of communication between \exes of different applications.

In a full-contention workload, as can be seen in Figure~\ref{fig:contention}(d),
\orexp similar to OX, executes the transactions sequentially, but,
because of the dependency graph generation overhead, its performance is a bit worse than OX.
The performance of the XOV paradigm, on the other hand, is highly reduced. Since all the transactions within a block conflict,
it can only commit one transaction per block (we reduced the block size of XOV to record its peak throughput).

In a full-contention workload with contention across applications (dashed line in Figure~\ref{fig:contention}(d)),
\orexp has high latency and low throughput.
Such a workload can be seen as a chain of translations where consecutive transactions belong to different applications.
As a result, to execute each transaction, a message exchange between a pair of \exes is needed.

 \input{graph_distance}

\vspace{-0.5em}
\subsection{Scalability over Multiple Data Centers}

In the last set of experiments, we measure the scalability of the blockchain systems over multiple data centers.
To this end, each time we move one group of nodes, i.e., clients, \ords, \exes, or non-executors,
to AWS Asia Pacific (Tokyo) Region data center, leaving the other nodes in the AWS US West Region data center.
We consider a no-contention workload.
The results can be seen in Figure~\ref{fig:distance}.

Moving the clients has the most impact on the XOV paradigm because in XOV clients participate in the first two phases.
Indeed, they send the requests to the \exes (endorsers), receive endorsements, and then send the endorsements to the orderer nodes.
Whereas in OX and OXII, clients send the requests and do not participate in other phases of the protocol.
As a result, as can be seen in Figure~\ref{fig:distance}(a), the delay of XOV becomes much larger.

Orderers are the core part of all three blockchains;
they receive transactions from clients, agree on the order of the transactions,
put the transactions into blocks, and send the blocks to every node.
As a result, moving them to a far data center, as shown in Figure~\ref{fig:distance}(b),
results in a considerable delay. Note that in OX, a subset of nodes are considered as orderers.

In the last two experiments, we move \exe (endorser) and non-exe nodes to the far data center.
Since there is no such a separation between nodes in the OX paradigm, we do not perform these two experiments.
Moving \exe nodes adds latency to the two phases of communication in XOV (clients to \exes and \exes to clients) and
one phase of communication in OXII (\ords to \exes). Note that when the \exes execute the messages and receive enough
number of matching results from other \exes, the transaction is counted as committed.
In addition, no communication between \exes is needed since we consider a no-contention workload.
Finally, moving non-executor nodes has no impact on the performance of OXII,
because the non-executor nodes are only informed about the blockchain state.
But in XOV, non-executors validate the blocks.
The results are shown in Figure~\ref{fig:distance}(c) and Figure~\ref{fig:distance}(d).
\vspace{-0.5em}

%% file: graph_block.tex
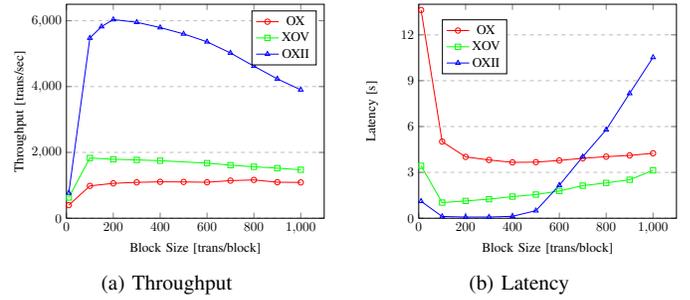
\begin{figure}[t!]
\begin{minipage}{.23\textwidth}\centering
\begin{tikzpicture}[scale=0.5]
\begin{axis}[
    xlabel={Block Size [trans/block]},
    ylabel={Throughput [trans/sec]},
    xmin=0, xmax=1100,
    ymin=0, ymax=6500,
    xtick={0,200,400,600,800,1000},
    ytick={0,2000,4000,6000},
    legend pos=north east,
    ymajorgrids=true,
    grid style=dashed,
]

\addplot[
    color=red,
    mark=o,
    ]
    coordinates {
    (10,398)(100,983)(200,1060)(300,1090)(400,1106)(500,1103)(600,1093)(700,1143)(800,1164)(900,1094)(1000,1088)};

\addplot[
    color=green,
    mark=square,
    ]
    coordinates {
    (10,630)(100,1830)(200,1791)(300,1772)(400,1748)(1717)(600,1674)(700,1612)(800,1563)(900,1521)(1000,1473)};

\addplot[
    color=blue,
    mark=triangle,
    ]
    coordinates {
    (10,770)(100,5476)(150,5823)(200,6036)(300,5953)(400,5793)(500,5600)(600,5360)(700,5020)(800,4620)(900,4233)(1000,3901)};

\addlegendentry{OX}
\addlegendentry{XOV}
\addlegendentry{OXII}

\end{axis}
\end{tikzpicture}
{\footnotesize (a) Throughput}
\end{minipage}\hfill
\begin{minipage}{.23\textwidth} \centering
\begin{tikzpicture}[scale=0.5]
\begin{axis}[
    xlabel={Block Size [trans/block]},
    ylabel={Latency [s]},
    xmin=0, xmax=1100,
    ymin=0, ymax=14,
    xtick={0,200,400,600,800,1000},
    ytick={0,3,6,9,12},
    legend style={at={(axis cs:100,13)},anchor=north west}, 
    ymajorgrids=true,
    grid style=dashed,
]

\addplot[
    color=red,
    mark=o,
    ]
    coordinates {
    (10,13.626)(100,5.014)(200,4.014)(300,3.813)(400,3.656)(500,3.671)(600,3.781)(700,3.921)(800,4.03)(900,4.11)(1000,4.250)};

\addplot[
    color=green,
    mark=square,
    ]
    coordinates {
    (10,3.431)(100,1.024)(200,1.13)(300,1.249)(400,1.414)(500,1.548)(600,1.782)(700,2.132)(800,2.311)(900,2.518)(1000,3.134)};

\addplot[
    color=blue,
    mark=triangle,
    ]
    coordinates {
    (10,1.109)(100,0.114)(200,0.078)(300,0.078)(400,0.126)(500,0.494)(600,2.160)(700,4.024)(800,5.790)(900,8.169)(1000,10.526)};

\addlegendentry{OX}
\addlegendentry{XOV}
\addlegendentry{OXII}

\end{axis}
\end{tikzpicture}
{\footnotesize (b) Latency}
\end{minipage}
\caption{Throughput/Latency Measurement by Increasing the Block Size}
  \label{fig:block}
  \vspace{-1em}
\end{figure}

%% file: graph_contention.tex
\begin{figure*}[t!]
\begin{minipage}{.23\textwidth}\centering
\begin{tikzpicture}[scale=0.49]
\begin{axis}[
    xlabel={Throughput [trans/sec]},
    ylabel={Latency [s]},
    xmin=0, xmax=6400,
    ymin=0, ymax=5,
    xtick={0,2000,4000,6000},
    ytick={1,2,3,4},
    legend style={at={(axis cs:2500,4.8)},anchor=north west}, 
    ymajorgrids=true,
    grid style=dashed,
]

\addplot[
    color=red,
    mark=o,
    ]
    coordinates {
    (140,1.390)(280,0.569)(660,0.523)(890,0.540)(1052,3.656)};

\addplot[
    color=green,
    mark=square,
    ]
    coordinates {
   (150,1.190)(500,0.493)(1400,0.570)(1800,0.613)(1860,1.24)(1980,4.231)
    };

\addplot[
    color=blue,
    mark=triangle,
    ]
    coordinates {
    (140,1.140)(666,0.281)(1306,0.161)(2493,0.110)(5880,0.072)(6036,0.078)(6060,4.702)};
    
\addplot[
    color=black,
    mark=+,
    dashed,
    ]
    coordinates {
    (140,1.140)(666,0.281)(1306,0.161)(2493,0.110)(5880,0.072)(6036,0.078)(6060,4.702)};

\addlegendentry{OX}
\addlegendentry{XOV}
\addlegendentry{OXII}
\addlegendentry{OXII*}

\end{axis}
\end{tikzpicture}
{\footnotesize (a) No-contention ($0\%$)}
\end{minipage}\hfill
\begin{minipage}{.23\textwidth} \centering
\begin{tikzpicture}[scale=0.49]
\begin{axis}[
    xlabel={Throughput [trans/sec]},
    ylabel={Latency [s]},
    xmin=0, xmax=6400,
    ymin=0, ymax=5,
    xtick={0,2000,4000,6000},
    ytick={1,2,3,4},
    legend style={at={(axis cs:2200,4.8)},anchor=north west}, 
    ymajorgrids=true,
    grid style=dashed,
]

\addplot[
    color=red,
    mark=o,
    ]
    coordinates {
    (140,1.390)(280,0.569)(660,0.523)(890,0.540)(1052,3.656)};

\addplot[
    color=green,
    mark=square,
    ]
    coordinates {
   (152,1.198)(403,0.503)(1170,0.592)(1413,0.643)(1590,1.142)(1680,4.311)
    };

\addplot[
    color=blue,
    mark=triangle,
    ]
    coordinates {
    (130,1.440)(502,0.383)(1086,0.202)(1993,0.140)(4323,0.231)(4702,0.406)(4930,3.271)};
    
\addplot[
    color=black,
    mark=+,
    dashed,
    ]
    coordinates {
    (121,1.642)(502,0.561)(1086,0.502)(1993,0.537)(3843,0.771)(4012,0.906)(4300,4.012)};

\addlegendentry{OX}
\addlegendentry{XOV}
\addlegendentry{OXII}
\addlegendentry{OXII*}

\end{axis}
\end{tikzpicture}
{\footnotesize (b) Low-contention ($20\%$)}
\end{minipage}\hfill
\begin{minipage}{.23\textwidth} \centering
\begin{tikzpicture}[scale=0.49]
\begin{axis}[
    xlabel={Throughput [trans/sec]},
    ylabel={Latency [s]},
    xmin=0, xmax=6400,
    ymin=0, ymax=5,
    xtick={0,2000,4000,6000},
    ytick={1,2,3,4},
    legend pos=north east,
    ymajorgrids=true,
    grid style=dashed,
]

\addplot[
    color=red,
    mark=o,
    ]
    coordinates {
    (140,1.390)(280,0.569)(660,0.523)(890,0.540)(1052,3.656)};

\addplot[
    color=green,
    mark=square,
    ]
    coordinates {
   (30,1.198)(261,0.592)(305,0.643)(379,1.142)(491,4.311)
    };

\addplot[
    color=blue,
    mark=triangle,
    ]
    coordinates {
    (190,0.843)(284,0.513)(410,0.469)(870,0.502)(1540,0.711)(1760,1.756)(1830,3.756)};
    
\addplot[
    color=black,
    mark=+,
    dashed,
    ]
    coordinates {
    (181,0.943)(245,0.583)(370,0.569)(806,0.713)(1103,1.111)(1213,1.82)(1272,2.156)(1280,4.256)};

\addlegendentry{OX}
\addlegendentry{XOV}
\addlegendentry{OXII}
\addlegendentry{OXII*}

\end{axis}
\end{tikzpicture}
{\footnotesize (c) High-contention ($80\%$)}
\end{minipage}\hfill
\begin{minipage}{.23\textwidth} \centering
\begin{tikzpicture}[scale=0.49]
\begin{axis}[
    xlabel={Throughput [trans/sec]},
    ylabel={Latency [s]},
    xmin=0, xmax=6400,
    ymin=0, ymax=5,
    xtick={0,2000,4000,6000},
    ytick={1,2,3,4},
    legend pos=north east,
    ymajorgrids=true,
    grid style=dashed,
]

\addplot[
    color=red,
    mark=o,
    ]
    coordinates {
    (140,1.390)(280,0.569)(660,0.523)(890,0.540)(1052,3.656)};

\addplot[
    color=green,
    mark=square,
    ]
    coordinates {
   (30,1.198)(170,0.592)(231,0.643)(281,1.142)(303,4.311)
    };

\addplot[
    color=blue,
    mark=triangle,
    ]
    coordinates {
    (103,1.602)(292,0.87)(582,0.793)(683,0.721)(793,0.763)(952,3.612)};
    
\addplot[
    color=black,
    mark=+,
    dashed,
    ]
    coordinates {
    (73,2.102)(232,0.97)(382,0.993)(483,1.021)(593,1.23)(692,4.012)};

\addlegendentry{OX}
\addlegendentry{XOV}
\addlegendentry{OXII}
\addlegendentry{OXII*}

\end{axis}
\end{tikzpicture}
{\footnotesize (d) Full-contention ($100\%$)}
\end{minipage}
\caption{Throughput/Latency Measurement by Increasing the Degree of Contention in the Workload}
  \label{fig:contention}
  \vspace{-1.5em}
\end{figure*}
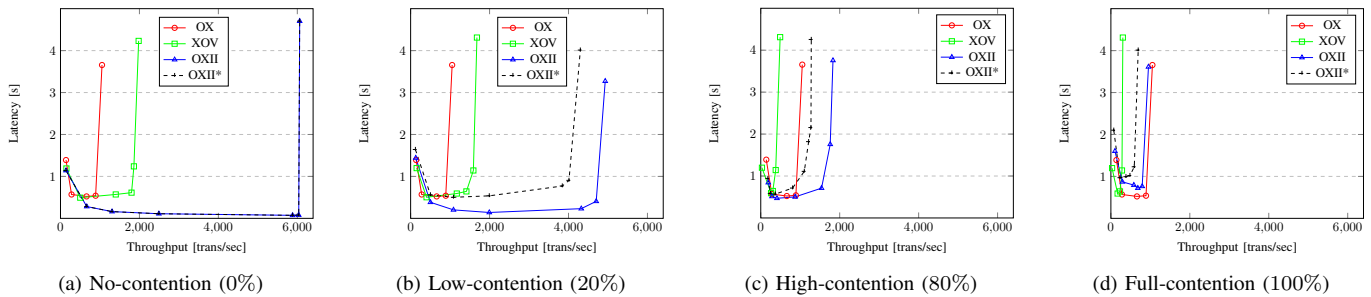

%% file: graph_distance.tex
\begin{figure*}[t!]
\begin{minipage}{.23\textwidth} \centering
\begin{tikzpicture}[scale=0.49]
\begin{axis}[
    xlabel={Throughput [trans/sec]},
    ylabel={Latency [s]},
    xmin=0, xmax=6400,
    ymin=0, ymax=6,
    xtick={0,2000,4000,6000},
    ytick={1,2,3,4,5},
    legend style={at={(axis cs:2500,5.8)},anchor=north west}, 
    ymajorgrids=true,
    grid style=dashed,
]

\addplot[
    color=red,
    mark=o,
    ]
    coordinates {
    (130,1.720)(240,0.901)(630,0.774)(810,0.840)(904,5.156)};

\addplot[
    color=green,
    mark=square,
    ]
    coordinates {
   (110,1.620)(380,0.983)(1210,1.053)(1370,1.313)(1430,1.74)(1490,5.731)
    };

\addplot[
    color=blue,
    mark=triangle,
    ]
    coordinates {
    (140,1.320)(640,0.471)(1276,0.371)(2310,0.344)(5230,0.297)(5490,0.342)(5537,5.302)};

\addlegendentry{OX}
\addlegendentry{XOV}
\addlegendentry{OXII}

\end{axis}
\end{tikzpicture}
{\footnotesize (a) Apart Client}
\end{minipage}\hfill
\begin{minipage}{.23\textwidth} \centering
\begin{tikzpicture}[scale=0.49]
\begin{axis}[
    xlabel={Throughput [trans/sec]},
    ylabel={Latency [s]},
    xmin=0, xmax=6400,
    ymin=0, ymax=6,
    xtick={0,2000,4000,6000},
    ytick={1,2,3,4,5},
    legend style={at={(axis cs:2500,5.8)},anchor=north west}, 
    ymajorgrids=true,
    grid style=dashed,
]

\addplot[
    color=red,
    mark=o,
    ]
    coordinates {
    (110,1.843)(211,1.01)(640,0.720)(710,1.040)(807,4.83)};

\addplot[
    color=green,
    mark=square,
    ]
    coordinates {
   (137,1.338)(404,0.593)(1270,0.870)(1490,0.913)(1682,1.59)(1708,5.231)
    };

\addplot[
    color=blue,
    mark=triangle,
    ]
    coordinates {
    (120,1.520)(640,0.471)(1276,0.401)(2210,0.382)(5030,0.257)(5090,0.411)(5137,4.602)};

\addlegendentry{OX}
\addlegendentry{XOV}
\addlegendentry{OXII}
 
\end{axis}
\end{tikzpicture}
{\footnotesize (b) Apart Orderers}
\end{minipage}\hfill
\begin{minipage}{.23\textwidth} \centering
\begin{tikzpicture}[scale=0.49]
\begin{axis}[
    xlabel={Throughput [trans/sec]},
    ylabel={Latency [s]},
    xmin=0, xmax=6400,
    ymin=0, ymax=6,
    xtick={0,2000,4000,6000},
    ytick={1,2,3,4,5},
    legend style={at={(axis cs:2500,5.8)},anchor=north west}, 
    ymajorgrids=true,
    grid style=dashed,
]

\addplot[
    color=green,
    mark=square,
    ]
    coordinates {
   (137,1.338)(604,0.793)(1270,0.870)(1490,0.913)(1682,1.59)(1798,5.731)
    };

\addplot[
    color=blue,
    mark=triangle,
    ]
    coordinates {
    (131,1.362)(623,0.419)(1287,0.306)(2334,0.250)(5612,0.202)(5811,0.223)(5821,5.02)};

\addlegendentry{XOV}
\addlegendentry{OXII}

\end{axis}
\end{tikzpicture}
{\footnotesize (C) Apart Executors}
\end{minipage}\hfill
\begin{minipage}{.23\textwidth} \centering
\begin{tikzpicture}[scale=0.49]
\begin{axis}[
    xlabel={Throughput [trans/sec]},
    ylabel={Latency [s]},
    xmin=0, xmax=6400,
    ymin=0, ymax=6,
    xtick={0,2000,4000,6000},
    ytick={1,2,3,4,5},
    legend style={at={(axis cs:2500,5.8)},anchor=north west}, 
    ymajorgrids=true,
    grid style=dashed,
]

\addplot[
    color=green,
    mark=square,
    ]
    coordinates {
   (161,1.329)(641,0.597)(1181,0.732)(1610,0.933)(1720,1.29)(1780,4.431)
    };

\addplot[
    color=blue,
    mark=triangle,
    ]
    coordinates {
    (140,1.140)(666,0.281)(1306,0.161)(2493,0.110)(5880,0.072)(6036,0.078)(6060,4.702)};

\addlegendentry{XOV}
\addlegendentry{OXII}
 
\end{axis}
\end{tikzpicture}
{\footnotesize (d) Apart Non-executors}
\end{minipage}
\caption{Throughput/Latency Measurement by Moving a Group of Nodes to a Further Data Center}
  \label{fig:distance}
  \vspace{-1.5em}
\end{figure*}
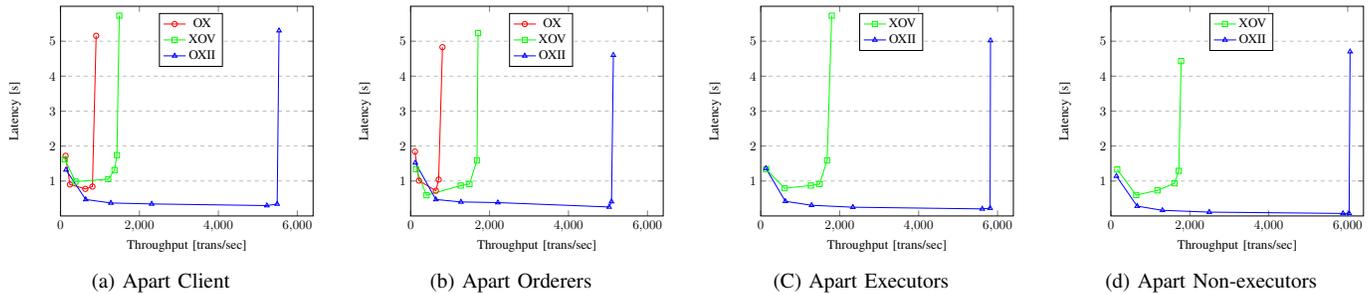

%% file: sec_related.tex
\section{Related Work}\label{sec:related}

The Order-execute paradigm is widely used in different permissioned blockchains.
Existing permissioned blockchains that employ the order-execute paradigm,
differ mainly in their ordering routines.
The ordering protocol of Tendermint \cite{kwon2014tendermint} differs from the original PBFT in two ways,
first, only a subset of nodes participate in the consensus protocol and second,
the leader is changed after the construction of every block (leader rotation).
Quorum \cite{morgan2016quorum} as an Ethereum-based \cite{ethereum17} permissioned blockchain
introduces a consensus protocol based on Raft \cite{ongaro2014search}: a well-known crash fault-tolerant protocol.
Chain Core \cite{chain}, Multichain \cite{greenspan2015multichain},
Hyperledger Iroha \cite{iroha}, and Corda \cite{corda} are some other prominent permissioned blockchains
that follow the order-execute paradigm.
As discussed in Section~\ref{sec:back}, these permissioned blockchains mainly suffer from
performance and confidentiality issues.
Hyperledger Fabric \cite{androulaki2018hyperledger} is a permissioned blockchain that employs
the execute-order(-validate) paradigm introduced by Eve \cite{kapritsos2012all}.
Fabric presents modular design, pluggable fault-tolerant protocol,
policy-based endorsement, and non-deterministic execution
for the first time in the context of permissioned blockchains.
Several recent studies attempt to improve the performance of Fabric
\cite{sousa2018byzantine,raman2018trusted,thakkar2018performance,fastfabric2019}.

We utilize some of the Fabric properties such as
modular design  and
pluggable fault-tolerant protocol
in \orexp.
However, \orexp is an {\em order-execute} paradigm.
In addition, while Fabric checks the read-write conflict in the last phase (validation)
which might result in transaction abort,
\orexp ensures correct results by generating dependency graphs in the first phase (ordering).
As a result, workloads with contention benefit most from \orexp.
Fabric also needs four phases of communications other than the ordering protocol
(clients to endorsers, endorsers to clients, clients to orderers, and orderers to peers)
while \orexp requires three phases
(clients to \ords, \ords to \exes, \exes to peers) which results in less latency.

BlockBench \cite{dinh2017blockbench} presents a framework for comparing the performance of
different blockchain platforms based on throughput, latency, scalability, and fault-tolerance.
In \cite{cachin2017blockchain} also a survey on some permissioned blockchain platforms
with respect to their consensus protocols is presented.

Our work is also related to concurrency control in DBMS.
Concurrency control is the activity of coordinating concurrent accesses to data \cite{bernstein1981concurrency}.
Concurrency control protocols mainly ensure the atomicity and isolation properties.
Many techniques have been proposed for concurrency control.
Lock-based protocols, e.g., two phase locking (2PL) \cite{bernstein1979formal,eswaran1976notions},
use locks to control the access to data.
Timestamp-based protocols \cite{bernstein1981concurrency,bernstein1987concurrency}
assign a global timestamp before processing where
by ordering the timestamp, the execution order of transactions is determined.
Optimistic Concurrency Control (OCC) \cite{kung1981optimistic} and Multi-Version Concurrency Control (MVCC) \cite{bernstein1983multiversion}
are two widely used timestamp-based protocols.
Dependency graphs are also, as discussed in section~\ref{sec:model},
used by several recent studies for concurrency control
\cite{faleiro2015rethinking, yao2015dgcc, faleiro2017high}.
\vspace{-0.5em}

%% file: sec_conc.tex
\section{Conclusion}\label{sec:conc}
In this paper, we proposed \orexp, an order-execute paradigm for
permissioned blockchain to support distributed applications that execute concurrently.
\orexp is able to handle the workload with conflicting transactions
without rolling back the processed transactions or executing transactions sequentially.
Conflicts between the transactions of a single application as well as
the transactions of different applications are addressed in \orexp.
We also presented \system, a permissioned blockchain system
designed specifically in the \orexp paradigm.
Our experimental evaluations show that in workloads with
conflicting transactions, \system 
shows a better performance in comparison to 
both order-execute and
execute-order permissioned blockchain systems.